\documentclass[a4paper]{jpconf}
\usepackage{graphicx}
\usepackage{amsmath}
\usepackage{hyperref}
\begin{document}
\title{Perfect discretization of path integrals\footnote{Based on a talk given at Loops'11, Madrid, May 27 \cite{talk_Madrid}}}

\author{Sebastian Steinhaus}

\address{Max Planck Institute for Gravitational Physics (Albert Einstein Institute), Am M\"uhlenberg 1, 14476 Golm, Germany}

\ead{sebastian.steinhaus@aei.mpg.de}

\begin{abstract}
In order to obtain a well-defined path integral one often employs discretizations. In the case of General Relativity these generically break diffeomorphism symmetry, which has severe consequences since these symmetries determine the dynamics of the corresponding system.

In this article we consider the path integral of reparametrization invariant systems as a toy example and present an improvement procedure for the discretized propagator. Fixed points and convergence of the procedure are discussed. Furthermore we show that a reparametrization invariant path integral implies discretization independence and acts as a projector onto physical states.
\end{abstract}

\section{Introduction and Motivation}

Discretizing a continuum theory in order to employ numerical treatment or to regularize path integrals has become an important tool. In addition to that, in some approaches to quantum gravity discrete structures are expected to be fundamental.

However, discretizations have several drawbacks, in particular if the continuous system contains symmetries, e.g. diffeomorphism symmetry in General Relativity, which are generically broken by the discretization \cite{Bahr:2009ku}. Furthermore, discretizations are not unique, since many different discretizations might lead to the same continuum physics\footnote{Actually extracting large scale physics from discrete quantum gravity approaches is one of the main problems.}. As soon as discrete structures are supposed to be fundamental, these ambiguities must be addressed.

Another important issue, in case the continuous symmetries are broken, are ``broken gauge'' modes, i.e. degrees of freedom which are considered to be gauge in the continuum. In the discrete setting these modes become physical, that is one has to integrate them out in a path integral approach. However, in the continuum limit, the symmetries get restored such that one sums over equivalent discretizations, which might lead to divergencies.

These issues can be overcome by implementing the symmetry of the continuous system into the discrete. In addition to restoring the continuous symmetry, the system is independent of the discretization and the discretization might be even unique \cite{Bahr:2011uj}. However, to achieve this one has to pull back the dynamics of the continuous system onto the lattice, i.e. to solve the dynamics. In case this is not possible, one can improve the discretization by a Wilsonian Renormalization procedure and implement the dynamics step by step.

Such a discretization is called a perfect discretization, for instance discussed in \cite{Hasenfratz:1993sp} for asymptotically free theories. This has been extended to systems with a diffeomorphism(-like) symmetry, e.g. reparametrization invariant systems \cite{Marsden}, Regge calculus without \cite{Rocek:1982fr} and with a cosmological constant \cite{Bahr:2009qc}. However, these approaches only discuss classical systems. In \cite{Bahr:2011uj} we have successfully extended the program of perfect discretizations to the quantized reparametrization invariant particle, the results will be presented here briefly.

\section{Reparametrization invariant system}

A classical system is described by its Lagrangian $L(q,\dot{q})$, which is a function of generalized coordinates $q(t)$ and its first time-derivative $\dot{q}(t)$, with the action given by $S=\int dt\, L(q,\dot{q})$. By introducing an auxiliary parameter $s$, the sytem is parametrised such that time $t$ becomes a variable itself. One can check that
\begin{equation} \label{eq:cl_act}
S\big(q(s),t(s)\big)= \int ds \, L\Big(q, \frac{q'}{t'}\Big) t'
\end{equation}
is invariant under reparametrizations, i.e. transformations of the form $q(s),\,t(s) \rightarrow q(f(s)),\, t(f(s))$. As a consequence, the solutions to the equations of motion to (\ref{eq:cl_act}) are not uniquely determined by the boundary data. In the canonical framework, this is manifest as the canonical momenta $p_q$ and $p_t$ (conjugated to $q$, $t$ respectively) are not linearly independent, which results in a constraint identical to the Hamiltonian $\tilde{H} = p_t + H = 0$ of the system, where $H$ denotes the Hamiltonian of the non-parametrized system.

To discretize the system, $q(s)$ and $t(s)$ are replaced by a finite set of variables, $q_n$ and $t_n$ with $n=0,\ldots,N$. The action is discretized by approximating derivatives by difference quotients and choosing a discretization of the potential term:
\begin{equation}
S^{(0)}(q_n,t_n) := \sum_{n=0}^{N-1} S^{(0)}_n := \sum_{n=0}^{N-1} \left[\frac{1}{2} \left(\frac{q_{n+1} -q_n}{t_{n+1} - t_n } \right)^2 + \frac{1}{2} \left(V(q_n) + V(q_{n+1}) \right) \right] (t_{n+t} - t_n)
\end{equation}
This is extended to a quantum system by defining the propagator:
\begin{equation}
K^{(0)}(q_n,t_n,q_{n+1},t_{n+1}) := \sqrt{\frac{1}{2 \pi \hbar (t_{n+1} - t_n)}} \exp\left\{- \frac{1}{\hbar} S^{(0)}_n \right\}
\end{equation}

\begin{figure}[h]
\begin{center}
\includegraphics{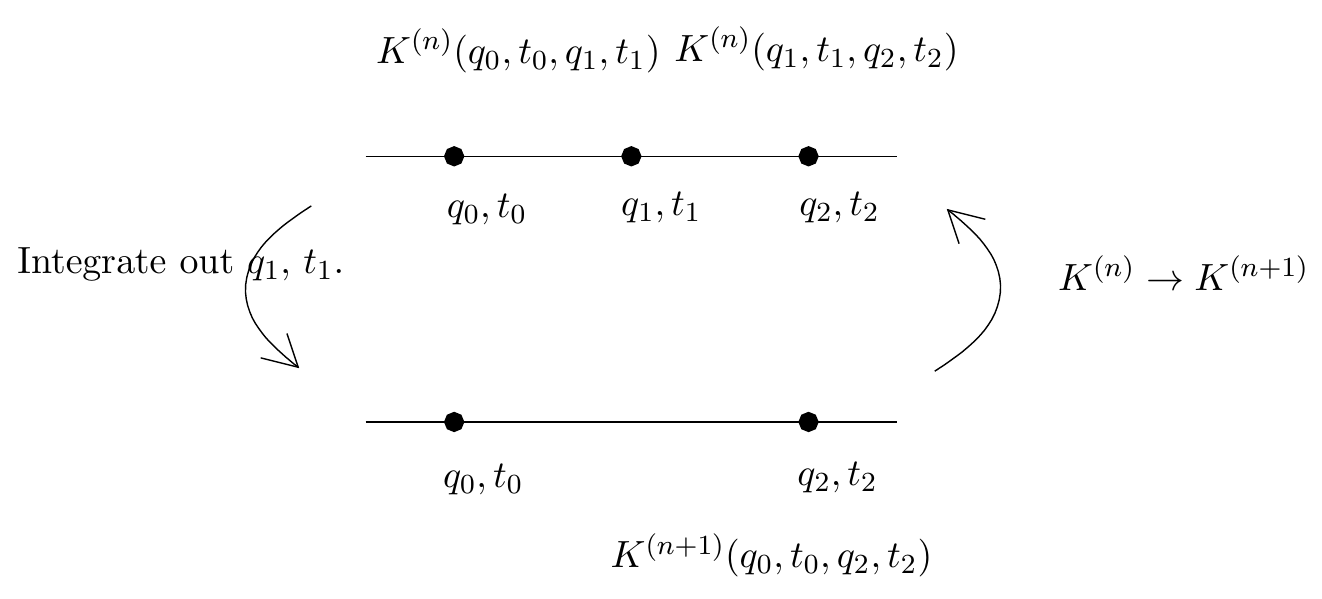}
\caption{By integrating out the variables $q_1$ and $t_1$, one obtains an improved propagator $K^{(n+1)}$.}
\label{fig:improve}
\end{center}
\end{figure}

In order to improve this discretization, consider a small discretization consisting of three lattice sites, where only the central lattice site with variables $q_1$ and $t_1$ is dynamical. The propagation between the lattice sites is described by $K^{(n)}(q_i,t_i,q_{i+1},t_{i+1})$, see also Fig. \ref{fig:improve}.
By integrating out $q_1$ and $t_1$, one obtains an improved propagator
\begin{equation} 
K^{(n+1)}(q_0,t_0,q_2,t_2) := \int dq_1 \, dt_1 \,K^{(n)}(q_0,t_0,q_1,t_1) K^{(n)}(q_1,t_1,q_2,t_2) . 
\end{equation}
Iterating this procedure gives the renormalization group flow of the system.

In order to discuss convergence of this procedure, we consider a class of systems by replacing the propagator by arbitrary coefficient functions. In case of the Harmonic oscillator, this gives:
\begin{equation} \label{eq:arb_prop}
K^{(n)}(q_0,q_1,T):=\eta^{(n)}(T) \, \exp\left\{-\frac{1}{\hbar} \left(\alpha^{(n)}_1 (T) (q_0^2 + q_1^2) + \alpha_2^{(n)} q_0 q_1 \right) \right\}
\end{equation}
Applying the discussed improving procedure to (\ref{eq:arb_prop}), one obtains recursion relations, e.g.
\begin{equation} \label{eq:rec_rel}
\alpha_1^{(n+1)} (2T) = \alpha^{(n)}_1 (T) - \frac{\alpha_2^{(n)}(T)^2}{8 \alpha_1(T)}\,.
\end{equation}
Fixed point equations can be obtained from (\ref{eq:rec_rel}) by dropping the superscripts. Solving these fixed point equations is equivalent to obtaining the continuum limit / solving the dynamics of the system. It turns out that the convergence of the initial system to the fixed point depends crucially on the choice of initial functions, in particular the measure. The procedure is convergent if
\begin{equation}
\eta^{(0)}(T) \sim \sqrt{\frac{1}{2 \pi \hbar T}}.
\end{equation} 
in leading order in $T$.

\section{Reparametrization invariance implies discretization independence}

So far we have not discussed the relation between the fixed point of the improvement procedure, i.e. a discretization independent system, and reparametrization invariance. 

Reparametrization invariance in the discrete is invariance of the action / path integral under vertex translations. Assume the propagator $K$ is invariant under vertex translations and consider the discretization described in Fig. \ref{fig:improve}. The path integral of this system is given by:
\begin{equation}
\langle q_0,t_0|q_2,t_2 \rangle := \int dq_1 \, dt_1 \, K(q_0,t_0,q_1,t_1) K(q_1,t_1,q_2,t_2)
\end{equation}
Since the $K$ is invariant under vertex translations, i.e. translations of $t_1$, we can gauge fix $t_1$ to an arbitrary value $t_1^f$ and drop the $t_1$ integration. Thus
\begin{equation}
\langle q_0,t_0|q_2,t_2\rangle = \int dq_1 \, K(q_0,t_0,q_1,t_1^f) K(q_1,t_1^f,q_2,t_2)
\end{equation}
Since $t_1^f$ is arbitrary, we can consider the limit $t_1^f \rightarrow t_2$. Since $K(q_1,t_1^f,q_2,t_2)$ gives the amplitude for a particle to propagate from $q_1$ to $q_2$ in the time interval $(t_2 - t_1^f)$, $K(q_1,t_1^f,q_2,t_2) \rightarrow \delta(q_2 - q_1)$ in the limit $t_1^f \rightarrow t_2$. But $t_1^f$ is arbitrary, such that we obtain:
\begin{equation} \label{eq:perf_prop}
K(q_0,t_0,q_2,t_2) = \int dq_1 K(q_0,t_0,q_1,t_1) K(q_1,t_1,q_2,t_2)
\end{equation}
Hence we have shown that reparametrization invariance realized in the discrete implies discretization independence of the system. Furthermore, (\ref{eq:perf_prop}) is the usual propagator property of quantum mechanics.

\section{Example: Anharmonic oscillator}

The same techniques allow to find the perfect discretization of the anharmonic oscillator with anharmonic term $\lambda q^4$, as a perturbation series in $\lambda$. For details, see \cite{Bahr:2011uj}.

One solution to the fixed point equations for the anharmonic oscillator to first order in the coupling constant $\lambda$ is given by:
\begin{align}
&K(q_0,q_1,T)=  \sqrt{\frac{\omega}{2 \pi \hbar \sinh(T \omega)}} \exp\left\{-\frac{1}{\hbar} S_{harm}(q_0,q_1,T)\right\} \times \Bigg(1 - \frac{\lambda}{\hbar} \times  \nonumber \\
 & \times \bigg[\frac{1}{768 \omega \sinh^4(\omega T)}\Big(12 T \omega - 8 \sinh(2 T \omega) + \sinh(4 T \omega)\Big) \left(q_0^4 + q_1^4\right) + \nonumber \\
 & + \frac{1}{192 \omega \sinh^4(T \omega)} \Big(-12 T \omega \cosh(T \omega) + 9 \sinh(T \omega) + \sinh(3 T \omega)\Big) \left(q_0^3 q_1 + q_0 q_1^3\right)+ \nonumber \\
 & + \frac{1}{64 \omega \sinh^4(T \omega)} \Big(2 T \omega (2 + \cosh(2 T \omega)) - 3 \sinh(2 T \omega)\Big) q_0^2 q_1^2 + \nonumber \\
 & + \hbar \left( \frac{2 + \cosh^2(T \omega) - 3 T \omega \coth(T \omega)}{32 \omega^2 \sinh^2(T \omega)}(q_0^2 + q_1^2) + \frac{4 T \omega+ 2 T \omega \cosh(2 T \omega) - 3 \sinh(2 T \omega)}{32 \omega^2 \sinh^3(T \omega)} q_0 q_1 \right)- \nonumber \\
 & - \frac{\hbar^2}{64 \omega^3} \left(3 \coth(T \omega) - T \omega \left(2 + \frac{3}{\sinh^2(T \omega)}\right)\right) \bigg] + O(\lambda^2)\Bigg) \label{eq:perf_prop_anh}
\end{align}
where $S_{harm}$ is the perfect action of the harmonic oscillator. Only the full expression (\ref{eq:perf_prop_anh}) satisfies the constraint equation (to first order in $\lambda$). For the interpretation and a more general solution of the fixed point equations, we refer to \cite{Bahr:2011uj}.

\section{Summary and Outlook}

To address the problem of broken gauge symmetries in discretized reparametrization invariant systems, we introduced the method of perfect discretizations. With this we can construct a reparametrization symmetry preserving discretization of the path integral as a fixed point of a renormalization flow. We have also proven that in these systems the issues of (broken) reparametrization invariance and discretization independence are equivalent. For the discussion of approximate solutions of the fixed point equations, uniqueness and (pseudo-)gauge modes, we refer to \cite{Bahr:2011uj}.

Applying coarse graining / to improve the discretization for quantum gravity models is left for future work. For some first steps, see \cite{Dittrich:2011zh}. Furthermore, in \cite{measure_Regge} we will discuss a measure for linearized Regge Calculus invariant under Pachner moves.

\end{document}